\documentclass{pasj00}

\begin{document}
\Received{2007/06/27}
\Accepted{2007/09/14}

\title{Hinode Observations of Vector Magnetic Field Change Associated
with a Flare on 2006 December 13}

\author{Masahito \textsc{Kubo},\altaffilmark{1,2}
Takaaki \textsc{Yokoyama},\altaffilmark{3}
Yukio \textsc{Katsukawa},\altaffilmark{4}
Bruce W \textsc{Lites},\altaffilmark{1}
Saku \textsc{Tsuneta},\altaffilmark{4}
Yoshinori \textsc{Suematsu},\altaffilmark{4}
Kiyoshi \textsc{Ichimoto},\altaffilmark{4}
Toshifumi \textsc{Shimizu},\altaffilmark{2}
Shin'ichi \textsc{Nagata},\altaffilmark{5}
Theodore D \textsc{Tarbell},\altaffilmark{6}
Richard A \textsc{Shine},\altaffilmark{6}
Alan M \textsc{Title},\altaffilmark{6}
and 
David \textsc{Elmore},\altaffilmark{1}}
 
\altaffiltext{1}{High Altitude Observatory, National Center for
Atmospheric Research\thanks{The National Center for Atmospheric Research
is sponsored by the National Science Foundation.}, P.O. Box 3000,
Boulder, CO 80307, United States} 
\email{kubo@ucar.edu}
\altaffiltext{2}{Institute of Space and Astronautical Science, Japan
Aerospace Exploration Agency, 3-1-1 Yoshinodai, Sagamihara, Kanagawa
229-8510}
\altaffiltext{3}{Department of Earth and Planetary Science, School of
Science, University of Tokyo, Hongo, Bunkyo-ku, Tokyo, 113-0033}
\altaffiltext{4}{National Astronomical Observatory of Japan, 2-21-1
Osawa, Mitaka, Tokyo 181-8588}
\altaffiltext{5}{Hida Observatory, Kyoto University, Kamitakara, Gifu 506-1314}
\altaffiltext{6}{Lockheed Martin Advanced Technology Center, O/ADBS, B/252 3251 Hanover Street, Palo Alto, CA 94304 United States}


%


\maketitle

\begin{abstract}
Continuous observations of a flare productive active region 10930 were
successfully carried out with the Solar Optical Telescope onboard the
Hinode spacecraft during 2007 December 6 to 19. 
We focus on the evolution of photospheric magnetic fields in this active
region, and magnetic field properties at the site of the X3.4 class flare,
using a time series of vector field maps with high spatial resolution.
The X3.4 class flare occurred on 2006 December 13 at the apparent 
collision site between the large, opposite polarity umbrae.
Elongated magnetic structures with alternatingly positive and negative
polarities resulting from flux emergence appeared one day before the 
flare in the collision site penumbra.
Subsequently, the polarity inversion
line at the collision site became very complicated.
The number of bright loops in Ca II H increased during the formation of these
elongated magnetic structures.
The flare ribbons and bright loops evolved along the polarity inversion line 
and one footpoint of the bright loop was located in a region having a large
departure of field azimuth angle with respect to its surroundings.
The SOT observations with high spatial resolution and high polarization
 precision reveal temporal change in fine structure of magnetic fields
 at the flare site:
some parts of the complicated polarity inversion line then
disappeared, and in those regions the azimuth angle of photospheric 
magnetic field changed by about 90 degrees, becoming more spatially
uniform within the collision site.
\end{abstract}

\section{Introduction}
What magnetic activity of the photosphere is responsible for
triggering solar flares? Many studies have attempted to address this
question through measurements of the
evolution of photospheric magnetic fields
associated with flares. 
Major flares most often occur around the sheared polarity inversion line in
complicated active regions.  Precision
measurements of the magnetic field vector with high spatial resolution are
necessary to reveal changes in its fine structure at the
flare site, and also to minimize the inference of 
spurious magnetic cancellation events. 
Moreover, continuous observations under stable conditions, and for long
duration are crucial for understanding formation of the sheared polarity
inversion line (i.e. for study of the energy storage process, see papers
cited in \cite{Priest2002}). 

The Hinode satellite (\cite{Kosugi2007}) is in a sun-synchronous polar
orbit so that it observes the Sun continuously for eight
months each year without interruption by spacecraft night. 
The Solar Optical Telescope (SOT, \cite{Tsuneta2007, Suematsu2007,
Ichimoto2007, Shimizu2007a}) onboard the Hinode is a unique facility
that has the capability of providing long sequence of vector magnetic 
field measurements at high spatial resolution.  Thus, a primary
objective of Hinode/SOT is to
obtain flare-related changes in the magnetic field under seeing-free
conditions. 
In this letter, we present an initial report on the evolution of an active
region 10930 producing 4 X-class flares and magnetic field, and we
discuss properties of the field and the chromospheric emission at the
flare site. 

\section{Observation and Data Analysis}
\begin{figure}
  \begin{center}
    \FigureFile(80mm,60mm){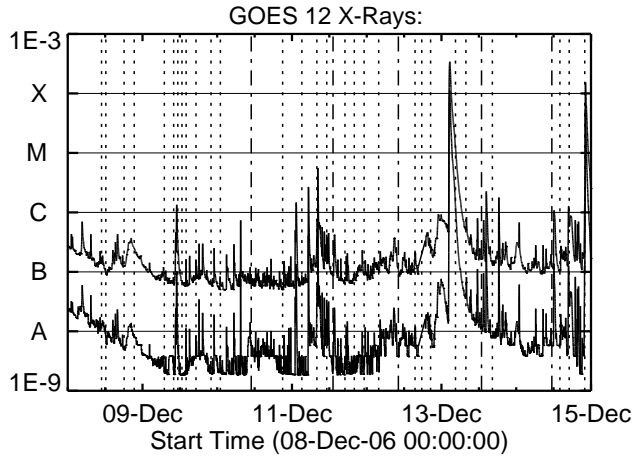}
  \end{center}
  \caption{Time profiles of the X-ray flux from the GOES 12
 satellite from 2007 December 8 to 15 are shown.
The upper and lower solid profiles indicate the full sun solar X-ray
 output in the 1-8 {\AA} and 0.5-4.0 {\AA} passbands, respectively. 
The vertical dotted lines (dot-dashed lines) represent the start time of
 the slit scanning for Fast Map (Normal Map) with the
 SOT/SP.}\label{fig:flare_goes}
\end{figure}

In December 2006, three X-class flares 
(X6.5 on December 6, X3.4 on December 13, and X1.5
on December 14) were observed with the SOT.  In this paper
we focus on the X3.4 flare on December 13, and on the associated
evolution of the
active region from December 8 to 15 while the active region was
located not too far from disk center.  For much of this period,
the SOT simultaneously obtained the following data:
the Broadband Filter Imager (BFI) provided G-band 4305 {\AA} and Ca II H
3968 {\AA} images, and the Narrowband Filter Imager (NFI) provided
longitudinal magnetograms in the photosphere. 
The G-band and Ca II H  are formed at temperatures of about 6000 K
(photosphere) and $10^4$K (chromosphere), respectively.
The time cadence of the BFI and NFI is 2 minutes.
The field of view was 218$\arcsec\times109\arcsec$ with pixel size of
0$\arcsec$.108 for the BFI and 327$\arcsec\times164\arcsec$ with pixel
size of 0$\arcsec$.16 for the NFI (these data are binned $2 \times 2$ pixels
from the full resolution).
The Spectro Polarimeter (SP) obtained the full polarization state
(I, Q, U, and V) of two magnetically sensitive Fe lines at 6301.5 {\AA} and
6302.5 {\AA} with the wavelength sampling of 21.6 m{\AA}.
The spatial distribution of magnetic field vectors over an active
region may be obtained by a scanning the solar image over the SP slit.
Figure \ref{fig:flare_goes} shows the GOES time variation of full Sun X-ray
flux. Also indicated on Figure \ref{fig:flare_goes} are the start times of 
SP mapping observations in two observing modes (Normal Map and Fast Map).
The spatial sampling was 0.$\arcsec$148$\times$0.$\arcsec$159 and
integration time for each slit position was 4.8s for the Normal Map mode.
The spatial sampling of the Fast Map mode was
0.$\arcsec$295$\times$0.$\arcsec$317 and the integration time was 3.2s.

The observation of full Stokes profiles permits us to derive 
accurate, quantitative measures of the magnetic field vector in the
photosphere.  
We used a non-linear least-squares fitting technique to fit analytical
Stokes profiles to the observed profiles (\cite{Yokoyama2007}).
In this study, we present the line-of-sight (LOS) magnetic field by
$B^{LOS}=Bf\cos(\gamma)$, where $B$ is the intrinsic strength of the
magnetic field, $f$ is filling factor, and $\gamma$ is the inclination
with respect to the LOS.
We have applied two methods for resolution of the 180$\degree$ azimuth
ambiguity: 
the AZAM utility \citep{Lites1993} for interactive resolution, and the
minimum energy method of \citet{Metcalf2006}. 
However, the azimuth angle ($\chi$) is presented as $\cos(2\chi)$ in
this paper. 
The measure $\cos(2\chi)$ is independent of the resolution of the
azimuth ambiguity. 
The SOT, for the first time, reveals sub-arcsecond fine structure
of the magnetic field vector around the polarity inversion line at the
flare site.
With resolution of this fine structure we encounter new difficulties in
resolution of the azimuth ambiguity.

Image co-alignment is necessary for comparison of images taken at
different times or at differing wavelengths. 
When we investigate the relationship between bright features
seen in Ca II H (using the BFI) and photospheric magnetic fields 
(using the SP), we determine their relative alignment
using an image cross-correlation technique.  We cross-correlate the
continuum intensity map from the SP with the G-band image taken at
the time of midpoint of the closest SP observation.
The relationship between centers of the field-of-view of the G-band
and Ca II H images is well known (\cite{Shimizu2007b}).
The continuum intensity from the SP maps taken at different times 
are also aligned by image cross-correlation in order to remove drifts due
to satellite jitter, correlation tracking, and sunspot proper motion.

\section{Results}

\subsection{Temporal Change of NOAA10930 for 7 Days}
\begin{figure*}
  \begin{center}
    \FigureFile(160mm,190mm){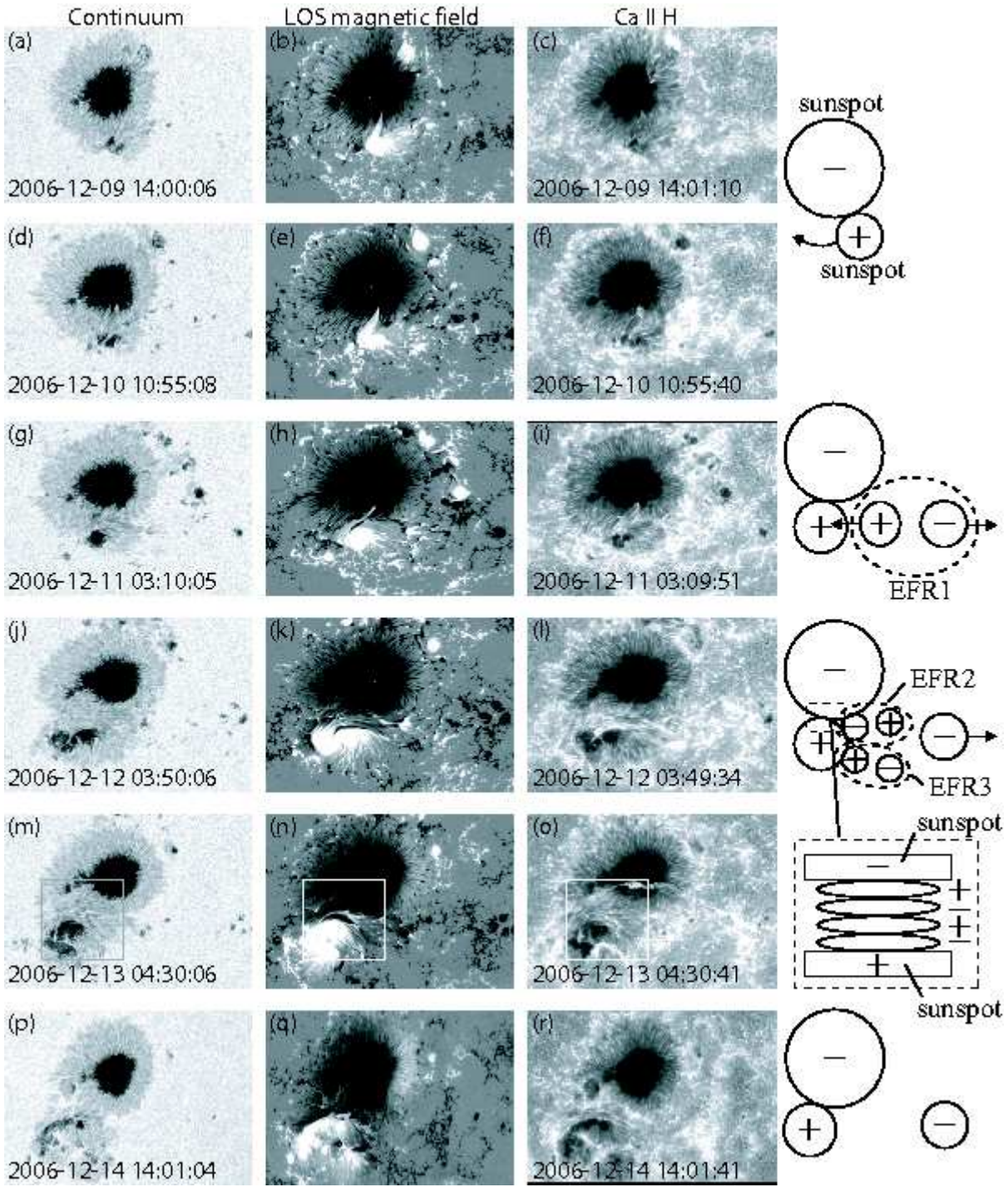}
  \end{center}
  \caption{Day-by-day evolution in NOAA10930.
Continuum intensity and LOS magnetic field maps are obtained
 with the SOT/SP. 
Ca II H images are observed with the SOT/BFI. North is up and
east is to the left.
The field of view is 128$\arcsec\times96\arcsec$.
The solid boxes indicate the field of view for Figures
 \ref{fig:flare_ribbon} and \ref{fig:flare_comp}.
}\label{fig:flare_1week}  
\end{figure*}

The active region NOAA10930 contained a $\delta$-sunspot comprising two
umbrae of opposite polarity in apparent ``collision'' 
(Figure \ref{fig:flare_1week}).
The southern umbra of positive polarity moved from west to east around
the large, mature northern umbra of negative polarity.
The positive umbra and its immediately surrounding penumbra
rotated counterclockwise.
An emerging flux region (EFR1 in the third row of Figure
\ref{fig:flare_1week}) appeared at the west side of the positive
umbra starting at about 9UT on December 10.
Then, the positive flux merged into the positive umbra, followed by
an increase in the size of the positive umbra.  
Another two flux emergence events (EFR2 and EFR3 in the forth row of Figure
\ref{fig:flare_1week}) began at about 0UT on 12 December. 
The emerging bipole of EFR3 had the same sense as EFR1 (positive-negative
oriented east-west), while the
emerging bipole of EFR2 was oriented in the sense opposite 
to the two other EFRs.
The negative flux of EFR2 merged into the penumbra surrounding the
southern positive umbra and followed a path around the positive umbra
that rotated counterclockwise.
As a result, elongated magnetic structures with alternating positive and
negative polarities were formed between the two sunspot umbrae.
Such elongated magnetic structures have been observed at other major
flare sites \citep{Zirin1993, Wang2005}.
The polarity inversion line formed at the boundary of at least 4
magnetic field systems (the two umbrae, EFR1, EFR2, and EFR3) became
very complicated before the flare.
The elongated magnetic structures gradually disappeared during the day
after the X3.4 flare (Figure \ref{fig:flare_1week} (q)), during which
time the
polarity inversion line also recovered its less complicated structure. 

Many bright loops in the Ca II H images were observed in the collision
site between the two umbrae.
In particular, the number of Ca II H bright loops suddenly increased
during the time when the negative flux of EFR2 entered into the region
around the positive umbra
(Figure \ref{fig:flare_1week} (l)).
When the X3.4 flare occurred on 2006 December 13, two flare ribbons in Ca
II H appeared in the collision site (Figure \ref{fig:flare_1week}
(o)).  
The duration of the two ribbon flare was about 6 hours.

\subsection{Photospheric Magnetic Field at the Flare Site}
\begin{figure*}
  \begin{center}
    \FigureFile(160mm,180mm){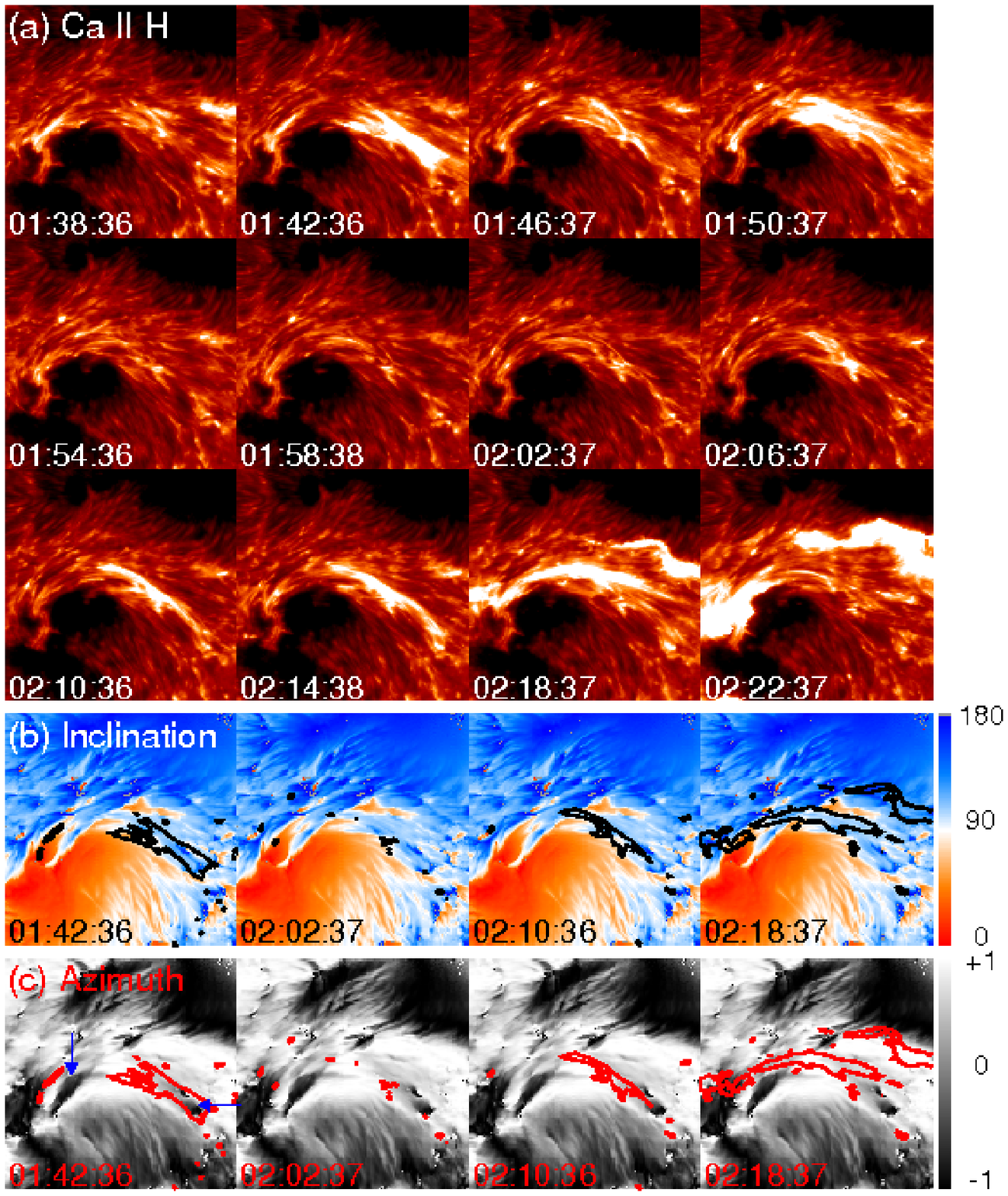}
  \end{center}
  \caption{(a) Time series of Ca II H images are observed with the
 SOT/BFI.
(b) Inclination ($\gamma$) map and (c) azimuth ($\chi$) map taken at
 20:30-21:33 on December 12 are made from the SOT/SP data.
The contours show the Ca II H intensity of 1500 DN.
The inclination represents the angle with respect to the
 LOS direction.
Inclination of 90 $\degree$ corresponds to magnetic field perpendicular to
 the LOS direction.
The azimuth map is represented by a definition of $\cos(2\chi)$.
The white area has magnetic field oriented east-west, while the
 black area has magnetic field oriented north-south.}
\label{fig:flare_ribbon} 
\end{figure*}

Figure \ref{fig:flare_ribbon} (a) shows the evolution of the two ribbon
flare observed in Ca II H.
The two ribbon flare starts from a small brightening at 02:02:37 around
the polarity inversion line in between the two sunspot umbrae (Figure
\ref{fig:flare_ribbon} (b)). 
Bright loops having a X-shape appear at 02:10:36, and then they evolve
into the southern ribbon of the flare.
The southern ribbon and other bright loops such as observed at 01:42:36
evolve along the polarity inversion line.
The two arrows in Figure \ref{fig:flare_ribbon} (c) indicate black
areas surrounded by white areas in the azimuth map.
This means that magnetic field in the black area is nearly perpendicular
to its surroundings.
The black areas are located at the polarity inversion line.
At least one side of the Ca II H bright loop is rooted at such a black area.

\begin{figure*}
  \begin{center}
    \FigureFile(160mm,100mm){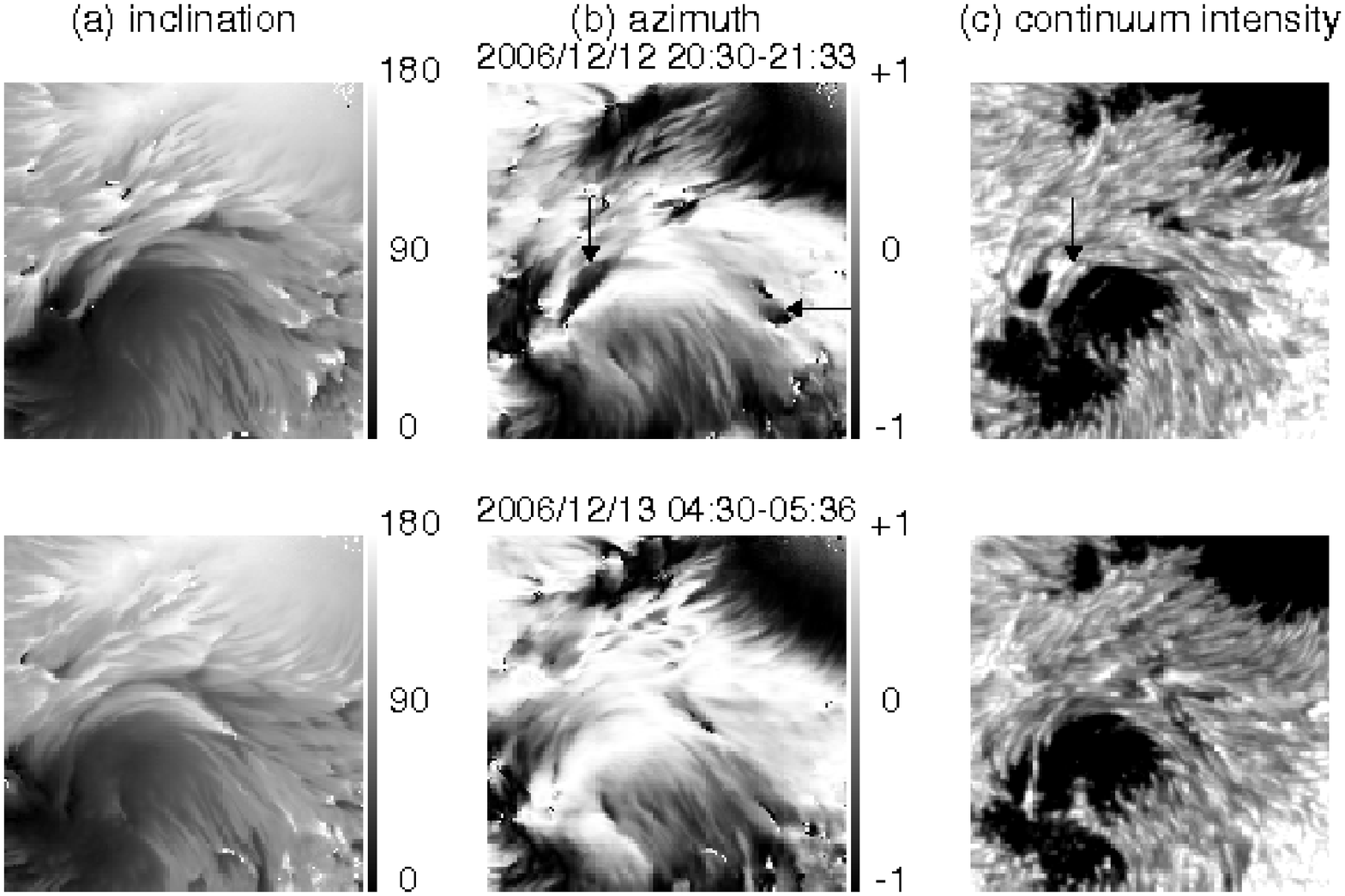}
  \end{center}
  \caption{(a) Inclination map, (b) azimuth map, and (c) continuum
 intensity map at the flare site. 
 The upper and lower panels are obtained at 20:30-21:33 on December 12
 and at 04:30 - 05:36 on December 13, respectively.
 }\label{fig:flare_comp}  
\end{figure*}

We examine temporal change in photospheric magnetic fields at the flare
site. 
The upper and lower panels of Figure \ref{fig:flare_comp} show magnetic fields
taken about 5 hours before and 3 hours after the flare onset, respectively.
Some of the complicated polarity inversion lines disappear after the
flare, and the polarity inversion line becomes relatively smooth overall
(Figures \ref{fig:flare_comp} (a)).
However, the magnetic field structure with alternating polarities still
remains between the two sunspots.
In the azimuth map, the areas with large azimuth angle with respect to
the surroundings (the two arrows in Figure \ref{fig:flare_comp} (b)) 
disappear after the flare: white (and gray) areas become dominant
around the polarity inversion line. 
This means that the azimuth angle changes by about 90$\degree$ in these areas.
The penumbra with negative polarity embedded in the southern positive
umbra disappears after the flare in one of the areas with large azimuth
angle with respect to the surroundings, as shown by the black arrow in
Figure \ref{fig:flare_comp} (c).

\section{Discussion}
Beginning one day before the X3.4 flare, many Ca II H bright loops
appear around the polarity inversion line between the opposite polarity
umbrae. 
Such pre-flare activity near the temperature minimum region has
been reported as brightening in the UV continuum observed with TRACE
(\cite{Kurokawa2002}). 
The SOT, for the first time, resolves the brightenings as 
apparent loops, and some of the bright loops have an X-shaped
configuration. 
The bright loops increase in number during the time that 
emerging negative elements migrate towards the positive umbra. 
Therefore, the bright loops are probably the result of magnetic
reconnection between magnetic field lines of the colliding opposite
polarity elements.

Many authors have shown that rapid and permanent changes of photospheric
magnetic fields are associated with large flares (e.g. \cite{Wang1992,
Cameron1999, Kosovichev2001, Spirock2002, Liu2005, Sudol2005, Wang2006}).
In this flare, there is one day gap between
the appearance of the elongated magnetic structures with
alternating polarities and the X-class flare on December 13. 
Such elongated magnetic structures are still observed during the day after
the flare. 
On the other hand, a change in shape of the polarity inversion line and 
a change in azimuth angle by 90$\degree$ are observed in small parts of the
polarity inversion line at the flare site during the 8 hour interval
between successive SP maps before and after onset of the flare.
This suggests that, in addition to
the variation in magnetic shear of the whole active region,
the temporal change in fine structure of magnetic fields at
the flare site would be related to the trigger mechanism
of the major flare.
\citet{Sudol2005} show that a typical timescale of the changes in
photospheric longitudinal magnetic fields is less than 10 minutes in
75$\%$ of 42 field change sites during 15 X-class flares. 
Observations of vector magnetic fields with higher cadence are needed
to reveal the actual magnetic activity responsible for 
triggering large solar flares.

We have found the change of azimuth angle by 90$\degree$ in some parts
of the polarity inversion line at the flare site.
However, there are two possibilities for such change as a result of the
azimuth ambiguity. 
One is that magnetic fields attain a similar direction to the surrounding
fields after the flare, and the other is that magnetic fields attain the
direction opposite to their surroundings.
The magnetic shear at the flare site decreases in the former case and
increases in the later case.
Moreover, the resolution of the azimuth ambiguity determines magnetic
field configuration of emerging bipoles (U-loop or $\Omega$-loop).
The global resolution of azimuth ambiguity may be determined by
several algorithms (\cite{Metcalf2006}), but it is very difficult to
resolve the ambiguity for each of the tiny magnetic elements resolved 
by the SOT. 
We need to resolve the azimuth ambiguity more carefully when calculating
the quantities relating to magnetic shear (e.g. magnetic shear angle, magnetic
helicity, current density, force-free alpha) from vector magnetic fields.

\bigskip
Hinode is a Japanese mission developed and launched by ISAS/JAXA, with
NAOJ as domestic partner and NASA and STFC (UK) as international
partners. It is operated by these agencies in co-operation with ESA and
NSC (Norway).
The authors would like to express their gratitude to late
Prof. T. Kosugi and all the members of Hinode team for a success of
the mission. 
This work was partly carried out at the NAOJ Hinode Science Center,
which is supported by the Grant-in-Aid for Creative Scientific Research
``The Basic Study of Space Weather Prediction from MEXT'', Japan (Head
Investigator: K. Shibata), generous donations from Sun Microsystems, and
NAOJ internal funding.

\end{document}